\documentclass[twocolumn,11pt]{article}
\usepackage{times}
\newdimen\proofrulebreadth \proofrulebreadth=.05em
\newdimen\proofdotseparation \proofdotseparation=1.25ex
\newdimen\proofrulebaseline \proofrulebaseline=2ex
\newcount\proofdotnumber \proofdotnumber=3
\let\then\relax
\def\hfi{\hskip0pt plus.0001fil}
\mathchardef\squigto="3A3B
\newif\ifinsideprooftree\insideprooftreefalse
\newif\ifonleftofproofrule\onleftofproofrulefalse
\newif\ifproofdots\proofdotsfalse
\newif\ifdoubleproof\doubleprooffalse
\let\wereinproofbit\relax
\newdimen\shortenproofleft
\newdimen\shortenproofright
\newdimen\proofbelowshift
\newbox\proofabove
\newbox\proofbelow
\newbox\proofrulename
\def\shiftproofbelow{\let\next\relax\afterassignment\setshiftproofbelow\dimen0 }
\def\shiftproofbelowneg{\def\next{\multiply\dimen0 by-1 }%
\afterassignment\setshiftproofbelow\dimen0 }
\def\setshiftproofbelow{\next\proofbelowshift=\dimen0 }
\def\setproofrulebreadth{\proofrulebreadth}

\def\prooftree{
\ifnum  \lastpenalty=1
\then   \unpenalty
\else   \onleftofproofrulefalse
\fi
\ifonleftofproofrule
\else   \ifinsideprooftree
        \then   \hskip.5em plus1fil
        \fi
\fi
\bgroup
\setbox\proofbelow=\hbox{}\setbox\proofrulename=\hbox{}%
\let\justifies\proofover\let\leadsto\proofoverdots\let\Justifies\proofoverdbl
\let\using\proofusing\let\[\prooftree
\ifinsideprooftree\let\]\endprooftree\fi
\proofdotsfalse\doubleprooffalse
\let\thickness\setproofrulebreadth
\let\shiftright\shiftproofbelow \let\shift\shiftproofbelow
\let\shiftleft\shiftproofbelowneg
\let\ifwasinsideprooftree\ifinsideprooftree
\insideprooftreetrue
\setbox\proofabove=\hbox\bgroup$\displaystyle 
\let\wereinproofbit\prooftree
\shortenproofleft=0pt \shortenproofright=0pt \proofbelowshift=0pt
\onleftofproofruletrue\penalty1
}

\def\eproofbit{
\ifx    \wereinproofbit\prooftree
\then   \ifcase \lastpenalty
        \then   \shortenproofright=0pt  
        \or     \unpenalty\hfil         
        \or     \unpenalty\unskip       
        \else   \shortenproofright=0pt  
        \fi
\fi
\global\dimen0=\shortenproofleft
\global\dimen1=\shortenproofright
\global\dimen2=\proofrulebreadth
\global\dimen3=\proofbelowshift
\global\dimen4=\proofdotseparation
\global\count255=\proofdotnumber
$\egroup  
\shortenproofleft=\dimen0
\shortenproofright=\dimen1
\proofrulebreadth=\dimen2
\proofbelowshift=\dimen3
\proofdotseparation=\dimen4
\proofdotnumber=\count255
}

\def\proofover{
\eproofbit 
\setbox\proofbelow=\hbox\bgroup 
\let\wereinproofbit\proofover
$\displaystyle
}%
\def\proofoverdbl{
\eproofbit 
\doubleprooftrue
\setbox\proofbelow=\hbox\bgroup 
\let\wereinproofbit\proofoverdbl
$\displaystyle
}%
\def\proofoverdots{
\eproofbit 
\proofdotstrue
\setbox\proofbelow=\hbox\bgroup 
\let\wereinproofbit\proofoverdots
$\displaystyle
}%
\def\proofusing{
\eproofbit 
\setbox\proofrulename=\hbox\bgroup 
\let\wereinproofbit\proofusing
\kern0.3em$
}

\def\endprooftree{
\eproofbit 
  \dimen5 =0pt
\dimen0=\wd\proofabove \advance\dimen0-\shortenproofleft
\advance\dimen0-\shortenproofright
\dimen1=.5\dimen0 \advance\dimen1-.5\wd\proofbelow
\dimen4=\dimen1
\advance\dimen1\proofbelowshift \advance\dimen4-\proofbelowshift
\ifdim  \dimen1<0pt
\then   \advance\shortenproofleft\dimen1
        \advance\dimen0-\dimen1
        \dimen1=0pt
        \ifdim  \shortenproofleft<0pt
        \then   \setbox\proofabove=\hbox{%
                        \kern-\shortenproofleft\unhbox\proofabove}%
                \shortenproofleft=0pt
        \fi
\fi
\ifdim  \dimen4<0pt
\then   \advance\shortenproofright\dimen4
        \advance\dimen0-\dimen4
        \dimen4=0pt
\fi
\ifdim  \shortenproofright<\wd\proofrulename
\then   \shortenproofright=\wd\proofrulename
\fi
\dimen2=\shortenproofleft \advance\dimen2 by\dimen1
\dimen3=\shortenproofright\advance\dimen3 by\dimen4
\ifproofdots
\then
        \dimen6=\shortenproofleft \advance\dimen6 .5\dimen0
        \setbox1=\vbox to\proofdotseparation{\vss\hbox{$\cdot$}\vss}%
        \setbox0=\hbox{%
                \advance\dimen6-.5\wd1
                \kern\dimen6
                $\vcenter to\proofdotnumber\proofdotseparation
                        {\leaders\box1\vfill}$%
                \unhbox\proofrulename}%
\else   \dimen6=\fontdimen22\the\textfont2 
        \dimen7=\dimen6
        \advance\dimen6by.5\proofrulebreadth
        \advance\dimen7by-.5\proofrulebreadth
        \setbox0=\hbox{%
                \kern\shortenproofleft
                \ifdoubleproof
                \then   \hbox to\dimen0{%
                        $\mathsurround0pt\mathord=\mkern-6mu%
                        \cleaders\hbox{$\mkern-2mu=\mkern-2mu$}\hfill
                        \mkern-6mu\mathord=$}%
                \else   \vrule height\dimen6 depth-\dimen7 width\dimen0
                \fi
                \unhbox\proofrulename}%
        \ht0=\dimen6 \dp0=-\dimen7
\fi
\let\doll\relax
\ifwasinsideprooftree
\then   \let\VBOX\vbox
\else   \ifmmode\else$\let\doll=$\fi
        \let\VBOX\vcenter
\fi
\VBOX   {\baselineskip\proofrulebaseline \lineskip.2ex
        \expandafter\lineskiplimit\ifproofdots0ex\else-0.6ex\fi
        \hbox   spread\dimen5   {\hfi\unhbox\proofabove\hfi}%
        \hbox{\box0}%
        \hbox   {\kern\dimen2 \box\proofbelow}}\doll%
\global\dimen2=\dimen2
\global\dimen3=\dimen3
\egroup 
\ifonleftofproofrule
\then   \shortenproofleft=\dimen2
\fi
\shortenproofright=\dimen3
\onleftofproofrulefalse
\ifinsideprooftree
\then   \hskip.5em plus 1fil \penalty2
\fi
}

\newtheorem{definition}{Definition}

\begin{document}
\global\def\refname{{\normalsize \it References:}}
\baselineskip 12.5pt
%
%
%
\title{\LARGE \bf Asynchronous Programming in a Prioritized Form}

\date{}


\author{\hspace*{-10pt}
\begin{minipage}[t]{3.6in} \normalsize \baselineskip 12.5pt
\centerline{Mohamed A. El-Zawawy$^{1,2}$} \centerline{$^{1}$College
of Computer and Information Sciences} \centerline{Al Imam Mohammad
Ibn Saud Islamic University } \centerline{Riyadh, Kingdom of Saudi
Arabia}  \centerline{} \centerline{$^{2}$Department of Mathematics}
\centerline{Faculty of Science} \centerline{Cairo
University}\centerline{Giza 12613, Egypt}
\centerline{maelzawawy@cu.edu.eg}
\end{minipage} \kern 0in
%
\\ \\ \hspace*{-10pt}
\begin{minipage}[b]{6.9in} \normalsize
\baselineskip 12.5pt {\it Abstract:} Asynchronous programming has
appeared as a programming style that overcomes undesired properties
of concurrent programming. Typically in asynchronous models of
programming, methods are posted into a post list for latter
execution. The order of method executions is serial, but
nondeterministic. \newline This paper presents a new and simple, yet
powerful, model for asynchronous programming. The proposed model
consists of two components; a context-free grammar and an
operational semantics. The model is supported by the ability to
express important applications. An advantage of our model over
related work is that the model simplifies the way posted methods are
assigned priorities. Another advantage is that the operational
semantics uses the simple concept of singly linked list to simulate
the prioritized process of methods posting and execution. The
simplicity and expressiveness make it relatively easy for analysis
algorithms to disclose the otherwise un-captured programming bugs in
asynchronous programs.
\\ [4mm] {\it Key--Words:}
Prioritized posting, asynchronous programming, operational
semantics, context-free grammar, concurrent programming.
\end{minipage}
\vspace{-10pt}}

\maketitle

\thispagestyle{empty} \pagestyle{empty}
%
%
\section{Introduction}
\label{S1} \vspace{-4pt}

All contemporary appliances (mobile, desktop, or web applications)
require high responsiveness which is conveniently provided by
asynchronous programming. Hence application program interfaces
(APIs) enabling asynchronous, non-blocking tasks, such as web access
or file operations) are accommodated in dominant programming
languages. APIs provide asynchronous programming but mostly in a
hard way. For example consider the following situation. A unique
user interface (UI) task thread is typically used to design and
implement user interfaces. Hence events on that thread simulate
tasks that change the UI state. Therefore when the UI  cannot be
redrawn or respond, it get freezed. This makes it sensible, in order
to keep the application responding continuously  to UI tasks, to run
blocking I/O commands and long-lasting CPU-bound asynchronously.

Asynchronous programming has multi-threaded roots. This is so as
APIs have been implemented using multi-threaded programs with
shared-memory. Software threads execution is not affected by the
number of processors in the system. This is justified by the fact
that the threads are executed as recursive sequential softwares
running concurrently with interleaved write and reads commands. The
many possible interleavings in this case cause the complexity of
models of concurrent programming~\cite{Kidd10}. In a complex
process, atomic locking commands can be added for prevention and
prediction of bad thread interleavings. The non-deterministic style
of interleaving occurrence creates rarely appearing
programming-errors which are typically very hard to simulate and
fix. This difficulty lead researchers to design multi-threaded
programs (of APIs) in the framework of asynchronous programming
models~\cite{Sen06,Jhala07}.

The relative simplicity of asynchronous programming makes it a
convenient choice to implement APIs or reactive systems. This is
proved by recent years intense use of asynchronous programming by
servers, desktop applications, and embedded systems. The idea of
asynchronous programming is to divide cumulative program executions
into tasks that are briefly-running. Moreover accessing the shared
memory, each task is executed as a recursive sequential software
that specifies (posts) new methods to be executed later.

Many attempts were made to present formal asynchronous programming
models. However few attempts~\cite{Emmi06} were made to express and
formalize the fact that posted tasks in asynchronous programs may
well have different execution priorities. A big disadvantage of
existing work~\cite{Emmi06} that considers execution priorities is
the complexity of the models hosting such priorities. For example
the work in~\cite{Emmi06} considers the execution priorities using
several task-buffers which makes the solution a bit involved.

This paper presents a simple, yet powerful, model for asynchronous
programming with priorities for task posting. We call the proposed
model $\mathcal{A}$synch$_\mathcal{P}$. The paper also presents a
novel and robust operational semantics for the constructs of
$\mathcal{A}$synch$_\mathcal{P}$. A simple singly linked list of
prioritized posted tasks is used to precisely capture the posting
process. Our proposed asynchronous model is to simplify analyzing
asynchronous programs~\cite{Atig08}.

\subsection*{Motivating Example}
\vspace{-4pt}

A motivating example of designing our model comes form the way
hardware interactions take place in operating systems (more
specifically in Windows) kernels. Concepts of prioritized interrupt
sets are used to simulate these hardware interactions in an
asynchronous style. For such applications a simple, yet powerful and
mathematically well founded, model for prioritized asynchronous
programming is required.

\subsection*{Contribution}
The contributions of this paper are:
\begin{itemize}
    \item A prioritized asynchronous programming model; $\mathcal{A}$synch$_\mathcal{P}$.
    \item A novel operational semantics for
    $\mathcal{A}$synch$_\mathcal{P}$ programs.
\end{itemize}

\subsection*{Organization}
The rest of the paper is organized as follows. Section~\ref{S2}
presents the proposed prioritized asynchronous programming model --
$\mathcal{A}$synch$_\mathcal{P}$. The semantics of prioritized
asynchronous programming model; $\mathcal{A}$synch$_\mathcal{P}$ is
shown in Section~\ref{S3} which is followed by Section~\ref{S4} that
reviews related work and presents directions for future work. The
last section (Section~\ref{S5}) of the paper concludes it.

\section{Prioritized Asynchronous Programming Model}
\label{S2} \vspace{-4pt}

\begin{figure*}[t]
\centering \fbox{
\begin{minipage}{12 cm}
{\footnotesize{
\begin{eqnarray*}
& & g\in G=\hbox{Global variable names.}\\
& & l\in L=\hbox{Method local variable names.}\\
& & p\in P=\{\hbox{high}(1),\hbox{medium}(2),\hbox{low}(3)\}\\
& & = \hbox{The set of all synchronization priorities.}\\
& & m\in M=\hbox{The set of all method names}.\\
& & v\in Val=\hbox{the set of possible values of local and global variables.} \\
S \in \hbox{Stats}   &::= &  S_1;S_2\mid g := e \mid l := e \mid
\hbox{Provided}\ e\mid  \hbox{if }e\ \hbox{then }S_t\ \hbox{else
}S_f\mid \\ & & \hbox{while }e\ \hbox{do }S\mid
   \hbox{run }m(e) \mid  \hbox{return}()\mid \hbox{Synch(m(e),p).}
\\ M\in \hbox{Meths}   &::= & \hbox{meth} (m,l,S).
\\
P\in \hbox{Programs}   &::= & \hbox{program} (g,M^*).
\end{eqnarray*}
}}
\end{minipage}
}\caption{$\mathcal{A}$synch$_\mathcal{P}$: a language model for a
simple prioritized asynchronous programming.}\label{f1}
\end{figure*}

This section presents our model, $\mathcal{A}$synch$_\mathcal{P}$,
for prioritized asynchronous programming. In
$\mathcal{A}$synch$_\mathcal{P}$, each posted method has an
execution priority. An asynchronous program execution is typically
divided into quick-running methods (tasks). Tasks of higher priority
get executed first and task of equal priorities are executed using
the first come first serviced strategy. Asynchronous programming has
an important application in reactive systems where a single task
must not be allowed to run too long and to prevent executing other
(potentially)  highly prioritized tasks.

Figure~\ref{f1} presents the simple and powerful model
$\mathcal{A}$synch$_\mathcal{P}$ for prioritized asynchronous
programming. Considering single local and global variables and using
free syntax of expressions does not cause any generality lose.
However each expression is built using the global variable of the
program and the local variable of the active method. A
$\mathcal{A}$synch$_\mathcal{P}$ program $P$ consists of a single
global variable $g$ and a sequence of methods denoted
$M_1,\ldots,M_n$.  The \textit{\hbox{Provided}(e)} statement
continues executing the program provided that the expression $e$
evaluates to a non-zero value.

Each method $M$ is expressed as a structure \textit{$\hbox{meth}
(m,l,S)$} of a method name, a single local variable $l$, and a
top-level statement $S$. The sets of all program methods and
statements are denoted by \textit{Meths} and \textit{Stmts},
respectively. Intuitively, the asynchronous call is modeled by the
statement $\hbox{Synch(m(e),p)}$ where:
\begin{itemize}
    \item the called method name is $m$,
    \item the calling parameter is the expression $e$, and
    \item the execution priority of this call is $p$.
\end{itemize}
We assume three levels of execution priorities;
$\{\hbox{high}(1),\hbox{medium}(2),\hbox{low}(3)\}$.

\section{Mathematical Framework for $\mathcal{A}$synch$_\mathcal{P}$}
\label{S3} \vspace{-4pt}

This section presents a novel operational semantics for asynchronous
programs built using $\mathcal{A}$synch$_\mathcal{P}$. Our semantics
is based on a singly liked-list (which we call \textit{Asynchronous
Linked List (ALL)}) to host the posted methods. \textit{ALL} is
divided into three regions using pointers. The first, the middle,
and last regions of \textit{ALL} host posted methods that have high,
medium, and low execution priorities, respectively.

Definition~\ref{D1} introduces formally the concept of
(\textit{Asynchronous Node (AN)}) to be used to build \textit{ALL}.
\begin{definition} \label{D1}
An asynchronous node (\textit{AN}), $n$, is a single linked list
node whose data contents are two locations containing:
\begin{itemize}
     \item $x_1$: a method name, and
     \item $x_2$: a parameter expression.
\end{itemize}
For a method call $m(e)$ in a $\mathcal{A}$synch$_\mathcal{P}$
program, we let Node$m(e)$ denotes the asynchronous node whose
locations $x_1 $ and $x_2$ contain $m$ and $e$, respectively. The
set of all asynchronous nodes is denoted by Nodes$_\mathcal{A}$.
\end{definition}

Definition~\ref{D2} introduces formally the concept of
\textit{Asynchronous Linked List (ALL)} that is to be used to
accurately capturing the semantics of the constructs of the proposed
asynchronous model.
\begin{definition} \label{D2}
An asynchronous linked list (ALL),
\begin{equation}li=<f,c,e_h,e_m>,\end{equation} is a singly
linked list whose nodes are asynchronous nodes (in
Nodes$_\mathcal{A}$) such that:
\begin{itemize}
    \item $f$ is a pointer to the first node of the list,
    \item $c$ is a pointer to the current node, and
    \item $e_h,e_m$ are pointers to the last node in the
    list hosting a method of high and medium priorities, respectively.
\end{itemize}
The set of all asynchronous linked lists is denoted by
Lists$_\mathcal{A}$.
\end{definition}

Whenever a method gets posted, an asynchronous node is created and
inserted into an asynchronous list. If the posted method is of
priority $h$ or $m$, the created node gets inserted after the nodes
pointed to by $e_h$ or $e_m$, respectively. If the posted method is
of priority $l$, the created node gets inserted at the end of the
list. Whenever a posted method is to be executed, the method
corresponding to the head of an asynchronous node is executed and
that head gets removed form the list. These two operations are
assumed to be carried out by the functions defined in
Definition~\ref{D3}.
\begin{definition} \label{D3}
Let $li=<f,c,e_h,e_m>$ be a asynchronous linked list (in
Lists$_\mathcal{A}$). We let
\begin{itemize}
    \item $\hbox{add}_\mathcal{A}: \hbox{Nodes}_\mathcal{A}\times P\times
    \hbox{Lists}_\mathcal{A}\rightarrow
    \hbox{Lists}_\mathcal{A}$ denotes a map that adds a given node $n$
    of a given priority $p$ after the node pointed to be $li.e_p$ in
    a given list $li$\footnote{Note that $p\in\{h,m,l\}$. If $p=l, e_l$
    is the last node in the list}.
    \item $\hbox{remove}_\mathcal{A}: \hbox{Lists}_\mathcal{A}\rightarrow
    \hbox{Nodes}_\mathcal{A}\times\hbox{Lists}_\mathcal{A}$ denotes a map that removes the first node of
    a given list $li$ and return the removed node and the resulting linked list.
\end{itemize}
\end{definition}

Definition~\ref{D4} introduces the states of our proposed
operational semantics.
\begin{definition} \label{D4}
Let program$(g,M_1,\ldots,M_n)$, where $M_i
=\hbox{meth}(m_i,l_i,S_i)$, be a program in
$\mathcal{A}$synch$_\mathcal{P}$. An asynchronous program state
(APS) is a triple (s,li,sk), where:
\begin{itemize}
    \item s is a partial map from $G\cup (M\times L)$ to Val.
    \item li is an asynchronous linked list.
    \item sk is stack of method names.
\end{itemize}
We let $M_i.l$ and $M_i.l$ denote $l_i$ and $S_i$, respectively.
\end{definition}
Each semantic state is a triple of a partial map captures the
contents of global and local variables, an asynchronous linked list,
and a stack of method names. The stack is meant to keep the order in
which methods call each other.

\begin{figure*}[t]
\centering \fbox{
\begin{minipage}{12cm}
{\footnotesize{
\[
\begin{prooftree}
li^\prime=\hbox{add}_\mathcal{A}(\hbox{Node}(m(e)),p,li) \justifies
\hbox{Synch}(m(e),p)):(s,li,sk)\rightarrow (s,li^\prime,sk)
\thickness=0.08em\using{(\hbox{synch}^s)}
\end{prooftree}
\]
\[
\begin{prooftree}
\hbox{is-empty}(sk)=\hbox{false}\qquad
sk^{\prime}=\hbox{pop}(sk)\justifies
\hbox{return}():(s,li,sk)\rightarrow (s,li,sk^{\prime})
\thickness=0.08em\using{(\hbox{return}^s)}
\end{prooftree}
\]
\[
\begin{prooftree}
\begin{tabular}{l}
$m^\prime=\hbox{peek}(sk)\qquad v=\|e(s(g),s(m^\prime.l)\|
 \qquad s^{\prime\prime}=s[m.l\mapsto v]$  \\
$sk^{\prime\prime}=\hbox{push}(sk,m)\qquad
m.S:(s^{\prime\prime},li,sk^{\prime\prime})\rightarrow
(s^\prime,li^\prime,sk^\prime)$
\end{tabular}
 \justifies \hbox{run
}m(e):(s,li,sk)\rightarrow (s^\prime,li^\prime,sk^\prime)
\thickness=0.08em\using{(\hbox{run}^s)}
\end{prooftree}
\]
\[
\begin{prooftree}
m^\prime=\hbox{peek}(sk)\qquad v=\|e(s.g,m^\prime.l)\| \justifies g
:= e:(s,li,sk)\rightarrow (s[g\mapsto v],li,sk)
\thickness=0.08em\using{(:=^s_g)}
\end{prooftree}
\]
\[
\begin{prooftree}
m^\prime=\hbox{peek}(sk)\qquad v=\|e(s.g,m^\prime.l)\|  \justifies l
:= e: (s,li,sk)\rightarrow (s[m^\prime.l\mapsto v],li,sk)
\thickness=0.08em\using{(:=^s_l)}
\end{prooftree}
\]
\[
\begin{prooftree}
m^\prime=\hbox{peek}(sk)\qquad \|e(s.g,m^\prime.l)\| \not = 0
\justifies \hbox{Provided}\ e:(s,li,sk)\rightarrow (s,li,sk)
\thickness=0.08em\using{(p^s)}
\end{prooftree}
\]
\[
\begin{prooftree}
m^\prime=\hbox{peek}(sk)\qquad \|e(s.g,m^\prime.l)\| = 0 \justifies
\hbox{while }e\ \hbox{do }S :(s,li,sk)\rightarrow
(s,li,sk)\thickness=0.08em\using{(w^s_1)}
\end{prooftree}
\]
\[
\begin{prooftree}
\begin{tabular}{l}
$m^\prime=\hbox{peek}(sk)\qquad \|e(s.g,m^\prime.l)\| \not = 0$\\
$S:(s,li,sk)\rightarrow(s^{\prime\prime},li^{\prime\prime},sk^{\prime\prime})
\qquad \hbox{while }e\ \hbox{do
}S:(s^{\prime\prime},li^{\prime\prime},sk^{\prime\prime})\rightarrow
(s^{\prime},li^{\prime},sk^{\prime}) $
\end{tabular}
\justifies \hbox{while }e\ \hbox{do }S :(s,li,sk)\rightarrow
(s^{\prime},li^{\prime},sk^{\prime})
\thickness=0.08em\using{(w^s_2)}
\end{prooftree}
\]
\[
\begin{prooftree}
\begin{tabular}{l}
$m^\prime=\hbox{peek}(sk)\qquad \|e(s.g,m^\prime.l)\|  = 0$\\
$S_f:(s,li,sk)\rightarrow(s^{\prime},li^{\prime},sk^{\prime})$
\end{tabular}
\justifies \hbox{if }e\ \hbox{then }S_t\ \hbox{else
}S_f:(s,li,sk)\rightarrow(s^{\prime},li^{\prime},sk^{\prime})
\thickness=0.08em\using{(\hbox{if}_1^s)}
\end{prooftree}
\]
\[
\begin{prooftree}
\begin{tabular}{l}
$m^\prime=\hbox{peek}(sk)\qquad \|e(s.g,m^\prime.l)\|  \not= 0$\\
$S_t:(s,li,sk)\rightarrow(s^{\prime},li^{\prime},sk^{\prime})$
\end{tabular}
\justifies \hbox{if }e\ \hbox{then }S_t\ \hbox{else
}S_f:(s,li,sk)\rightarrow(s^{\prime},li^{\prime},sk^{\prime})
\thickness=0.08em\using{(\hbox{if}_2^s)}
\end{prooftree}
\]
\[
\begin{prooftree}
S_1:(s,li,sk)\rightarrow
(s^{\prime\prime},li^{\prime\prime},sk^{\prime\prime}) \quad
S_2:(s^{\prime\prime},li^{\prime\prime},sk^{\prime\prime})\rightarrow
(s^{\prime},li^{\prime},sk^{\prime}) \justifies
S_1;S_2:(s,li,sk)\rightarrow (s^{\prime},li^{\prime},sk^{\prime})
\thickness=0.08em\using{(;^s)}
\end{prooftree}
\]
}}
\end{minipage}
}\caption{Transition rules for statements.}\label{f4}
\end{figure*}

\begin{figure*}[t]
\centering \fbox{
\begin{minipage}{12cm}
{\footnotesize{
\[
\begin{prooftree}
\justifies(s,\hbox{empty},\hbox{empty})\Rightarrow
(s,\hbox{empty},\hbox{empty})
 \thickness=0.08em\using{(\Rightarrow_1^s)}
\end{prooftree}
\]\[
\begin{prooftree}
\begin{tabular}{l}
$(n,li^\prime)=\hbox{remove}_\mathcal{A}(li)\qquad
n.x_1.S:(s,li^{\prime},\hbox{empty})\rightarrow
(s^{\prime\prime},li^{\prime\prime},\hbox{empty})$ \\
$(s^{\prime\prime},li^{\prime\prime},\hbox{empty})\Rightarrow
(s^\prime,\hbox{empty},\hbox{empty})$
\end{tabular}
\justifies (s,li,\hbox{empty})\Rightarrow
(s^\prime,\hbox{empty},\hbox{empty})
\thickness=0.08em\using{(\Rightarrow_2^s)}
\end{prooftree}
\]
\[
\begin{prooftree}
sk^{\prime\prime}=\hbox{push}(m,sk)\qquad
S:(s,li,sk^{\prime\prime})\rightarrow(s^{\prime},li^{\prime},sk^{\prime})
\justifies \hbox{meth}(m,l,S):
(s,li,sk)\rightarrow(s^{\prime},li^{\prime},sk^{\prime})
\thickness=0.08em\using{(\hbox{meth}^s)}
\end{prooftree}
\]
\[
\begin{prooftree}
\begin{tabular}{l}
$\forall 1\le i\le n.\ M_i:(s_i,li_i,sk_i)\rightarrow
(s_{i+1},li_{i+1},sk_{i+1})$   \\
$(s_{n+1},li_{n+1},\hbox{empty})\Rightarrow
(s^\prime,\hbox{empty},\hbox{empty})$
\end{tabular} \justifies
\hbox{program}(g,M_1,\ldots,M_n): (s_1,li_1,sk_1)\rightarrow
(s^\prime,\hbox{empty},\hbox{empty})
\thickness=0.08em\using{(\hbox{prog}^s)}
\end{prooftree}
\]
}}
\end{minipage}
}\caption{Transition rules for methods and programs.}\label{f5}
\end{figure*}

Figures~\ref{f4} and~\ref{f5} present the transition rules of the
proposed operational semantics. Some comments on the rules are in
order. The rule $\hbox{synch}^s$ creates an asynchronous node
corresponding to the method $m$ and the parameter $e$. Using the map
$\hbox{add}_\mathcal{A}$, the node then is added to the asynchronous
list $li$ to get the new list $li^\prime$. The rule
$(\hbox{return}^s)$, pops an element from the method stack as the
\textit{return} statement means that the top element of the stack is
executed. The rule $(\hbox{run}^s)$ first peeks the first element of
the stack to get the local variable ($m^\prime.l$) of the currently
active method. This local variable is then used together with the
global variable to evaluate the expression $e$. The resulting value
is used to modify the local variable of the  method ($m$) that is to
be executed. Then $m$ is pushed into the stack and the statement of
$m$ is executed.

The rule $(\hbox{prog}^s)$ first runs the statements of all methods
of the program being executed then runs all statements of the
methods that is posted. The posted statements are executed via the
rules $(\Rightarrow_1^s)$ and $(\Rightarrow_2^s)$.

\section{Related and Future Work}
\label{S4} \vspace{-4pt}

Parallel, distributed, reactive, and concurrent programming have
been attracting much researcher activities. The asynchronous
programming methodologies include:  \begin{itemize}
    \item multi-threaded light-weight orchestration programming~\cite{Ranjan12},
    \item thread Join-based allocation,
    \item typed synchronous programming languages~\cite{Aguado14},
    \item functional sensible programming,
    \item promises, and
    \item co-methods and futures agents~\cite{Gori14,Nelson12}.
\end{itemize}
Event-based techniques for programming have been using continuations
which are delimited monadic~\cite{Holzer11,Vaseux13}. Fork-join,
task, async, and event functions appear not to rely on a specific
language design. There is a big research debate about the
relationship between threads and events in systems
research~\cite{Kerneis11}.

In an asynchronous program, executions containing context switches
bounded by a user-specified limit are explored by context-bounded
verification~\cite{Qadeer04,Qadeer05}. This context-bounding idea is
not reasonable for programs with big number of events. Several
treatments for this problem have been proposed for this problem.
Without losing decidability,~\cite{Atig09} proposed a
context-minimizing technique permitting unbounded context switches.
For asynchronous concurrent programs, in~\cite{Torre10}, the
round-robin technique for scheduling is used to enable unbounded
context switches.

Sequential techniques are also used to analyze  asynchronous
programs. In~\cite{Qadeer04}, a source-to-source technique building
sequential programs from multi-threaded programs was proposed via
under approximating  the possible set of executions of the input
program. A novel source-to-source transformation providing for any
context-bound, a context-bounded under approximation was presented
in~\cite{Lal09}. A main issue of the work in~\cite{Lal09} is that
the resulting sequential program may host main states unreachable in
the given asynchronous program. Other techniques like~\cite{Torre09}
treat this problem by repeatedly running the code to the control
points where used-defined valued are needed. The work
in~\cite{Ghafari10} compares the techniques of asynchronous programs
verifications that use verification-condition-checking against that
use model-checking. One major results of this work is that eager
approaches outperforms lazy ones. The work in~\cite{Kidd10} uses the
construction using a bound on the task number, to reduce
asynchronous programs into sequential programs via
priority-preemptive schedulers.

The work presented in this paper is close to
sequentialization~\cite{Qadeer04}; the concept describing
compositional reductions to get sequential programs from concurrent
ones. Although sequentialization started by checking multi-threaded
programs with one context-switch, it was developed later to treat a
user-specified number of context-switches. These switches occur
among statically-specified group of threads running using RR
order~\cite{Lal09}. In~\cite{Emmi11}, a technique for treating
context switches among an unspecified number of dynamically-created
tasks was presented. This technique (in~\cite{Emmi11}) hence
explicitly treats event-oriented asynchronous programs.

For future work, it is interesting to devise static analyses for
asynchronous programs using the model
$\mathcal{A}$synch$_\mathcal{P}$~\cite{El-Zawawy14}. Initial
experiments show that our proposed model is expected to support
devising robust and powerful analysis techniques. An examples of
targeted analyses is \textit{dead-posting elimination} which aims at
removing the unnecessary posting statements from asynchronous
programs.

\section{Conclusion}
\label{S5} \vspace{-4pt} Main reason to use asynchronous programming
is to overcome some problems of concurrent programming. The main
idea of asynchronous programming is to post methods into a post list
for latter execution. The order of executing these methods is
nondeterministically serial. \newline A new and simple, yet
powerful, model for asynchronous programming was presented in this
paper. More precisely, the paper proposed a context-free grammar and
an operational semantics for asynchronous programming. One important
aspect of the proposed model is supporting posting methods with
execution priorities.

\end{document}